\documentclass[fleqn,usenatbib,useAMS]{mnras}

\usepackage{graphicx}	
\usepackage{amsmath}	
\usepackage{amssymb}	
\usepackage{multicol}        
\usepackage{bm}		
\usepackage{pdflscape}	
\usepackage[T1]{fontenc}

\usepackage{natbib}
\usepackage{hyperref}
\usepackage[normalem]{ulem}
\usepackage{lscape}
\usepackage{mathtools}
\usepackage{color}
\usepackage{soul}
\usepackage{longtable}
\usepackage{ulem}
\usepackage{makecell}
\usepackage{multirow}

\usepackage{graphicx}
\usepackage{txfonts}

\title[Energy levels and transitions for Ce~III]{Theoretical investigation of energy levels and transitions for Ce~III with applications to kilonova spectra}

   \author[G. Gaigalas et al.]
	 {G. Gaigalas,$^{1}$\thanks{E-mail: gediminas.gaigalas@tfai.vu.lt}
		P. Rynkun,$^{1}$
    N. Domoto,$^{2}$
    M. Tanaka,$^{2,3}$
    D. Kato$^{4,5}$
    and L. Kitovien\.{e}$^{1}$
		\\
	$^1$Institute of Theoretical Physics and Astronomy, Vilnius University, Saul\.{e}tekio Ave. 3, LT-10257 Vilnius, Lithuania\\
	$^2$Astronomical Institute, Tohoku University, Sendai 980-8578, Japan
\\	
	$^3$Division for the Establishment of Frontier Sciences, Organization for Advanced Studies, Tohoku University, Sendai 980-8577, Japan
\\	
	$^4$National Institute for Fusion Science, 322-6 Oroshi-cho, Toki 509-5292, Japan\\
	$^5$Interdisciplinary Graduate School of Engineering Sciences, Kyushu University, Kasuga, Fukuoka 816-8580, Japan}

\date{Accepted XXX. Received YYY; in original form ZZZ}

\pubyear{2024}

\begin{document}
\label{firstpage}
\pagerange{\pageref{firstpage}--\pageref{lastpage}}
\maketitle
 
\begin{abstract}
Doubly ionized cerium (Ce$^{2+}$) is one of the most important 
ions to understand the kilonova spectra.
In particular, near-infrared (NIR) transitions of Ce III between the ground (5p$^6$ 4f$^2$) and first excited (5p$^6$ 4f 5d) configurations are responsible for the absorption features around 14,500 \AA. 
However, there is no dedicated theoretical studies to provide accurate transition probabilities for these transitions.
We present energy levels of the ground and first excited configurations and transition data between them for Ce III. 
Calculations are performed using the {\sc Grasp}2018
package, which is based on the multiconfiguration Dirac–Hartree–Fock and relativistic configuration interaction methods. 
Compared with the energy levels in the NIST database, our calculations reach the accuracy with the root-mean-square (rms) of 
2732 cm$^{-1}$ or 1404 cm$^{-1}$ (excluding one highest level) for ground configuration, and rms of 618 cm$^{-1}$ for the first excited configuration.
We extensively study the line strengths and find that the Babushkin gauge provide the more accurate values.
By using the calculated $gf$ values, we show that the NIR spectral features of kilonova can be explained by the Ce III lines.
\end{abstract}


\section{Introduction}

Binary neutron star (NS) mergers have been thought to be the origin of rapid neutron capture ($r$-process) elements.
In 2017, the gravitational wave (GW) from the NS merger (GW170817) was successfully detected, and the associated electromagnetic counterpart (AT2017gfo) was observed \citep{Abbott17}.
The observed properties of AT2017gfo in UV/optical/near-infrared (NIR) wavelengths were found to be consistent with what expected as a kilonova \citep[e.g.,][]{Kasen17, Tanaka17, Kawaguchi18, Rosswog18}, thermal emission powered by radioactive decay of $r$-process nuclei \citep[e.g.,][]{Li98,Metzger10,Tanaka16,Metzger19}.
This fact provided us with evidence that $r$-process nucleosyntehsis occurs in NS mergers.

To interpret the observational properties of kilonovae, atomic data for heavy elements synthesized in NS merger ejecta are necessary.
In NS merger ejecta, photons interact with matter mainly via bound-bound transitions before escaping from the system.
Bound-bound opacity plays a major role to determine the behavior of kilonova light curves \citep[e.g.,][]{Kasen13,Tanaka13}.
This requires the complete knowledge of the bound-bound transitions for all the heavy elements.
Since such data are not available experimentally,
theoretical atomic calculations for heavy elements have been performed to construct the atomic data \citep[e.g.,][]{Kasen13, Kasen17, Gaigalas_2019_Nd, Gaigalas_2020_Er, Gaigalas_2022_Pr, Tanaka18, Tanaka20, Fontes20, Banerjee20, Banerjee22}.
Such theoretical data have been used to evaluate the opacities in kilonova ejecta, which provides the foundation of light curve modeling of kilonovae.

On the other hand, 
to interpret the detailed spectral features of kilonovae, 
more accurate atomic data are necessary.
\citet{Domoto_2022} have pointed out the importance of doubly ionized Ce ($Z$ = 58) in the spectra of kilonovae.
The Ce III transitions produce distinct absorption features in the spectra at NIR wavelengths:
three transitions at $\sim$ 16000 {\AA} between energy levels of $\mathrm{5p^64f^2-5p^64f\,5d}$ cause the features.
In fact, the features caused by Ce III nicely match with the observed features in the spectra of AT2017gfo.
However, since the transition probabilities ($gf$-values) of the transitions are experimentally unknown, they adopted the theoretical values \citep{Tanaka20} whose accuracy is not certain.
\citet{Domoto_2023} further tested the identification of Ce III in the spectra of AT2017gfo, by estimating the $gf$-values of the three Ce III lines using absorption lines in stellar spectra.
Such, so-called ``astrophysical $gf$-values'' broadly agree with those used in \citet{Domoto_2022}, which currently support the identification of Ce III in the spectra of AT2017gfo.

In fact, spectra of Ce$^{2+}$ ion have been studied by several experiments.
\citet{Sugar_1965} performed observations of the spectrum of Ce III from 757 to 11 091 \AA\,
 and 
discovered one hundred twenty-six newly energy levels, including revised values of previously known levels.
\citet{Johansson_1972} also measured the wavelengths of the $\mathrm{5p^64f^2-5p^64f\,5d}$ transition in a region 
between 11000 - 26 000 \AA.
These data of the Ce$^{2+}$ ion were critically evaluated by \citet{Martin_crit_eval}
and are given in the Atomic Spectra Database (ASD) of the National Institute of Standards 
and Technology (NIST, \citealt{NIST_ASD}).

\citet{Andersen_1974} measured the lifetimes for six excited levels (of the $\mathrm{5p^64f\,6p}$ configuration) using the beam-foil method.
Radiative lifetimes of nine levels of the $\mathrm{5p^64f\,6p}$ configurations were measured by \citet{Li_2000_PRA} 
using the time-resolved laser-induced fluorescence technique.
They also presented transition probabilities, these were obtained from
branching fractions calculated by the Cowan code \citep{Cowan} and the experimental lifetimes.

Atomic data of this ion have also been studied with semi-empirical and {\em ab-initio} theoretical calculations. 
Mainly the ground configuration or low excited configurations were investigated.
\citet{Bord_1997_MNRAS} used the Cowan code to calculate oscillator strengths.
\citet{Wyart_1998} investigated the energy levels and transition parameters using parametric fit, and 
reported ten new levels and classified more than 70 new lines.
\citet{Biemont_2002_MNRAS} studied the importance of core-polarization effects on oscillator strength in Ce III using the relativistic Hartree–Fock (HFR) method.
\citet{QUINET_2004_ADNDT} used the HFR method to calculate the Land{\'e} $g$-factors for doubly ionized lanthanides ($Z$=57-71). 
\citet{Stanek_2004} used the multiconfiguration Dirac-Fock method to study transition parameters for $\mathrm{6s^2~^1S_0-6s\,6p~^1P_1,^3P_1}$ transitions in rare earth ionized systems (from La$^{+1}$ through Nd$^{+4}$).
\citet{Li_2014} used semi-empirical methods to compute M1 and E2 transition data 
within the ground configurations of some Ba-like and Dy-like ions.
\citet{Safronova_2015_PRA} used a configuration interaction approach
with second-order perturbation theory and a linearized coupled-cluster all-order method to compute excitation energies of 
the levels of ground and first excited configurations.
\citet{Fischer_2019_PRA} investigated the effect of electron correlation
on the energy levels of the ground configuration [Xe]$\mathrm{4f^2}$ using the {\sc Grasp} code.
\citet{Gallego_2021_MNRAS} used the relativistic multiconfiguration Dirac-Hartree-Fock to compute
energy spectra and radiative transition data.

In this paper, we perform {\it ab-initio} calculations for Ce$^{2+}$ with the {\sc Grasp}2018 \citep{grasp2018} code by focusing on the ground ($\mathrm{5p^64f^2}$) and first excited ($\mathrm{5p^64f\,5d}$) configurations.
Then, toward the application to NIR spectral feature of kilonovae, we compute electric dipole (E1) transitions.
This paper is organized as follows. In Section \ref{sec:grasp18}, we describe our computational procedures. We show our results in Section \ref{sec:results}, by providing intensive evaluation of energy level data and transition data. 
In Section \ref{sec:spec}, we apply our atomic data to the kilonova spectra and discuss the impact to the NIR spectral features. 
Finally, we give conclusions in Section \ref{sec:conclusions}.


\section{Computational procedure and scheme}
\label{sec:grasp18}
The calculations are performed using the {\sc Grasp}2018 package, which  
is based on the multiconfiguration Dirac-Hartree-Fock (MCDHF) 
and relativistic configuration interaction (RCI) methods.
More details about these methods can be found in \citet{topical_rev} and \citet{grant}.

In the paper, only the main steps of the computational procedure are described.
The initial wave functions were generated in the same way as in the previous computations of lanthanide ions
\citep{Gaigalas_2019_Nd,Gaigalas_2020_Er,Radziute_2020,Radziute_2021,Rynkun_2021_Ce,Gaigalas_2022_Pr}.
At first, the MCDHF computations of the ground [Xe]$\mathrm{4f^2}$ configuration were performed.
These orbitals were kept frozen and used for the next steps of the computations.
Further, the $\mathrm{5d}$ orbitals belonging to the first excited configuration set were optimized.

The calculations for even and odd states were performed simultaneously.
In the next steps of the MCDHF computation, active spaces (AS) of CSFs were generated by allowing single-double (SD) electron substitutions from the $\mathrm{5s, 5p, 4f, 5d}$ shells to the orbital spaces (OS): $\mathrm{OS_1 = \{6s,6p,6d,5f,5g\}}$, $\mathrm{OS_2 = \{7s,7p,7d,6f,6g,6h\}}$, 
$\mathrm{OS_3 = \{8s,8p,8d,7f,7g,7h\}}$. 
When a new OS is computed, the previous orbitals are frozen.
The [Kr]$\mathrm{4d^{10}}$ defines an inactive closed core and no substitutions were allowed from it.
The MCDHF calculations were performed in the extended optimal level (EOL) scheme \citep{EOL}.
Table \ref{summary} summarizes the calculations performed
for even and odd configurations by showing their $J$ and parity values and the
ASFs that were included in the optimization process.
Based on the orbitals from the MCDHF calculations, further RCI calculations, 
including the Breit interaction and leading QED effects, were performed.

RCI calculations were performed in the extended CSFs basis. 
The electron correlations were extended by opening closed $\mathrm{3d, 4s, 4p, 4d}$ shells step by step to find the most appropriate computational scheme for energies and transition parameters calculations.
The intermediate results are described and discussed in Section \ref{sec:results}.
The scheme chosen for the final calculations include electron correlations when 
SD substitutions were allowed from the $\mathrm{4d, 5s, 5p}$ 
and $\mathrm{4f, 5d}$ shells to the $\mathrm{OS_3}$, 
single-restricted double (SrD) substitutions were allowed from the $\mathrm{4s, 4p}$ shells to the $\mathrm{OS_1}$ (\textbf{+4d SrD4s4p} scheme). 
The restrictions on substitutions were applied, while allowing substitutions from the deeper core shells increases the AS very rapidly.
The number of CSFs in the final even and odd state expansions
distributed over the different $J$ symmetries is 8975504 and 14265384 for even and odd parity, respectively.

\begin{table}
\caption{Summary of computed eigenvalues for each $J$ of the even and odd configurations in extended optimal level scheme.}
\label{summary}
\centering
\begin{tabular}{c c c cc c c c}
\hline\hline
\multicolumn{1}{c}{$J$} & \multicolumn{1}{c}{Parity}&\multicolumn{1}{c}{Eigenvalues} &&
\multicolumn{1}{c}{$J$} & \multicolumn{1}{c}{Parity}&\multicolumn{1}{c}{Eigenvalues} \\
\hline
\noalign{\smallskip}
0  &+&   1-2 &&    0   &$-$& 1  \\
1  &+&   1   &&    1   &$-$& 1-3  \\
2  &+&   1-3 &&    2   &$-$& 1-4 \\
3  &+&   1   &&    3   &$-$& 1-4 \\
4  &+&   1-3 &&    4   &$-$& 1-4 \\
5  &+&   1   &&    5   &$-$& 1-3 \\
6  &+&   1-2 &&    6   &$-$& 1  \\
\hline
\hline
\end{tabular}
\end{table}

\section{Results}
\label{sec:results}
The accuracy of the computed results was evaluated comparing with data from NIST ASD \citep{NIST_ASD} 
and other theoretical computations.
In the following sections the influence of the correlations to the energy levels and transition data was studied.
The most appropriate computational scheme was chosen for both (energies and transition parameters) calculations.

\subsection{Evaluation of energy spectra}
\label{Evaluation_energies}

\begin{figure}
\centering
\includegraphics[width=\hsize]{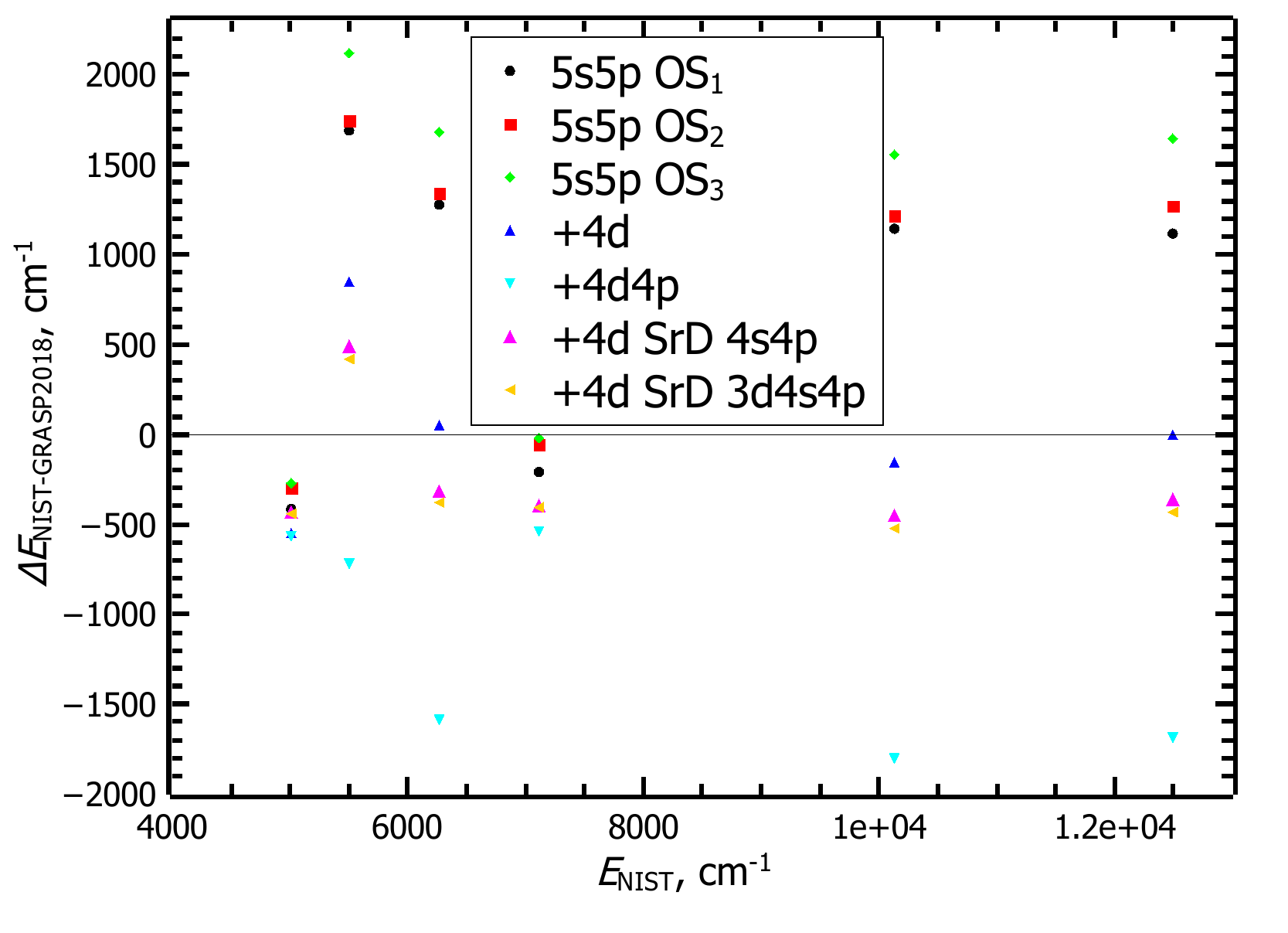}
\caption{\label{strategy_comparison} Differences between NIST ASD energy levels and those of the present {\sc Grasp}2018 calculations using different computational schemes (in cm$^{-1}$). }
\end{figure}

The influence of the correlations was studied by opening the closed shells step by step for substitutions.
To reduce the computational resources, the importance of the correlations of the closed shells was studied for the levels 
of the ground configurations with $J$ = 4 (since it is the ground level) 
and the levels of the first excited configuration with $J$ = 3. 
The contributions of the correlations are presented in Fig. \ref{strategy_comparison}. 
The figure shows that correlations from the 4d shell are important.
By opening the 4p shell the energies of the ground configuration almost do not change; 
the energies of the first excited configuration are too high.
Restricting the substitutions from the 4p and 4s shells (allowing only S substitutions; \textbf{+4d SrD4s4p} scheme) improves the agreement with the NIST data.
When the 3d shell is opened, the agreement remains similar.
The importance of correlations for transition data was also studied and described in Section \ref{Evaluation_transition}.

\begin{figure}
\centering
\includegraphics[width=\hsize]{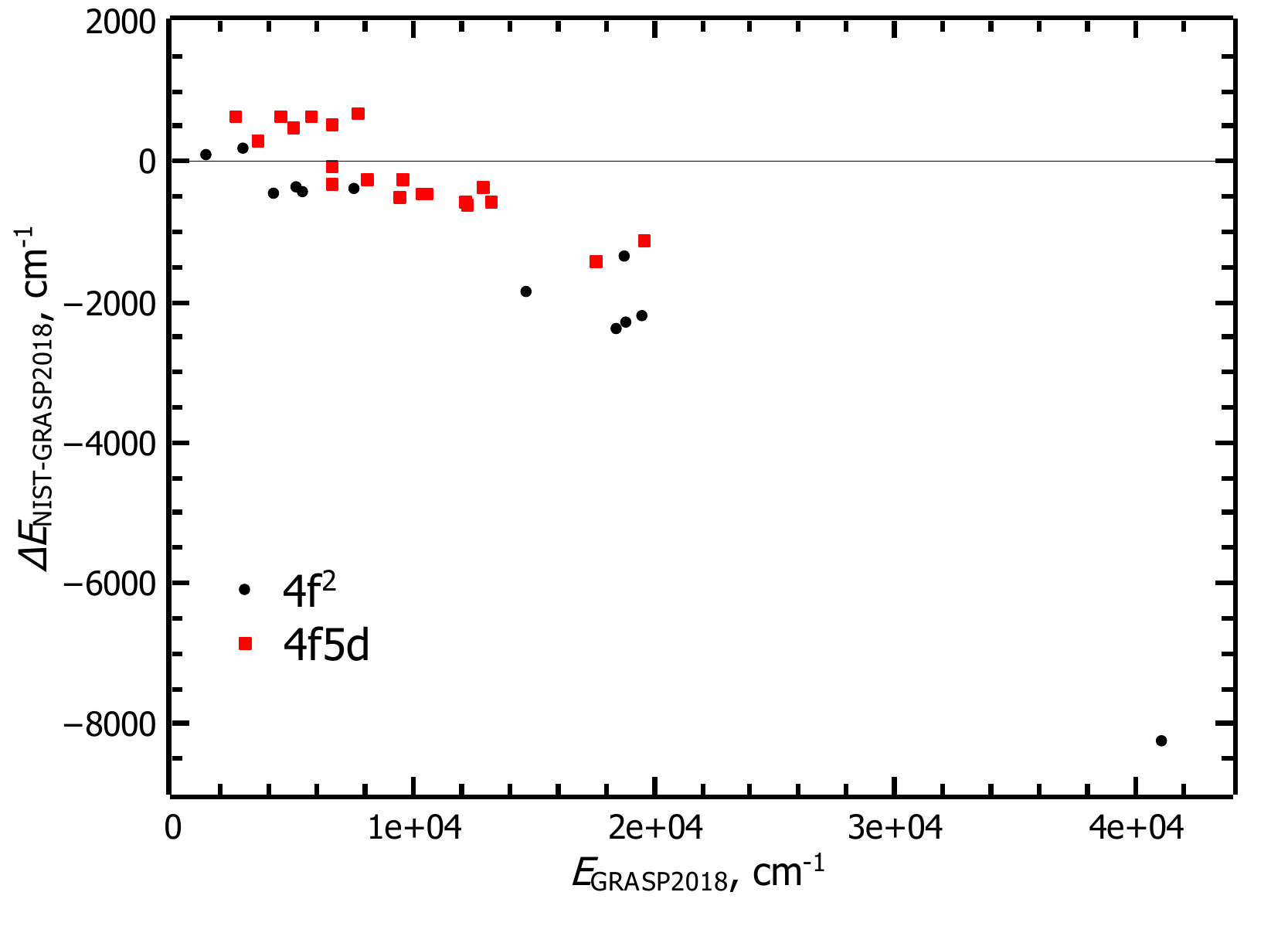}
\caption{\label{comparison_NIST_final_results} Differences between NIST ASD energy levels and those of the present {\sc Grasp}2018 calculations
(in cm$^{-1}$).}
\end{figure}

Therefore, the \textbf{+4d SrD4s4p} scheme was chosen to calculate the energy levels for both configurations. 
Fig. \ref{comparison_NIST_final_results} presents the comparison of the final results (using  \textbf{+4d SrD4s4p} scheme) with the NIST ASD.
As seen in the figure, the differences for the energy levels of two configurations up to 12000 cm$^{-1}$ energy reaches 600 cm$^{-1}$.
The disagreement of other energies reaches 2500 cm$^{-1}$, and the largest difference (8200 cm$^{-1}$) is for the level of the ground configuration (4f$^2~^1S_0$). The level 4f$^2~^1S_0$ is remote from the other levels of the ground configuration, its energy is 41076 cm$^{-1}$. Therefore, other additional correlations are relevant for this level.
The root-mean-square (rms) deviations obtained for the energy levels of the ground configuration from the NIST data are 2732 cm$^{-1}$, but excluding the level with the worst disagreement (4f$^2~^1S_0$), the rms is 1404 cm$^{-1}$. The rms for the first excited configuration is 618 cm$^{-1}$.

\begin{figure}
\centering
\includegraphics[width=\hsize]{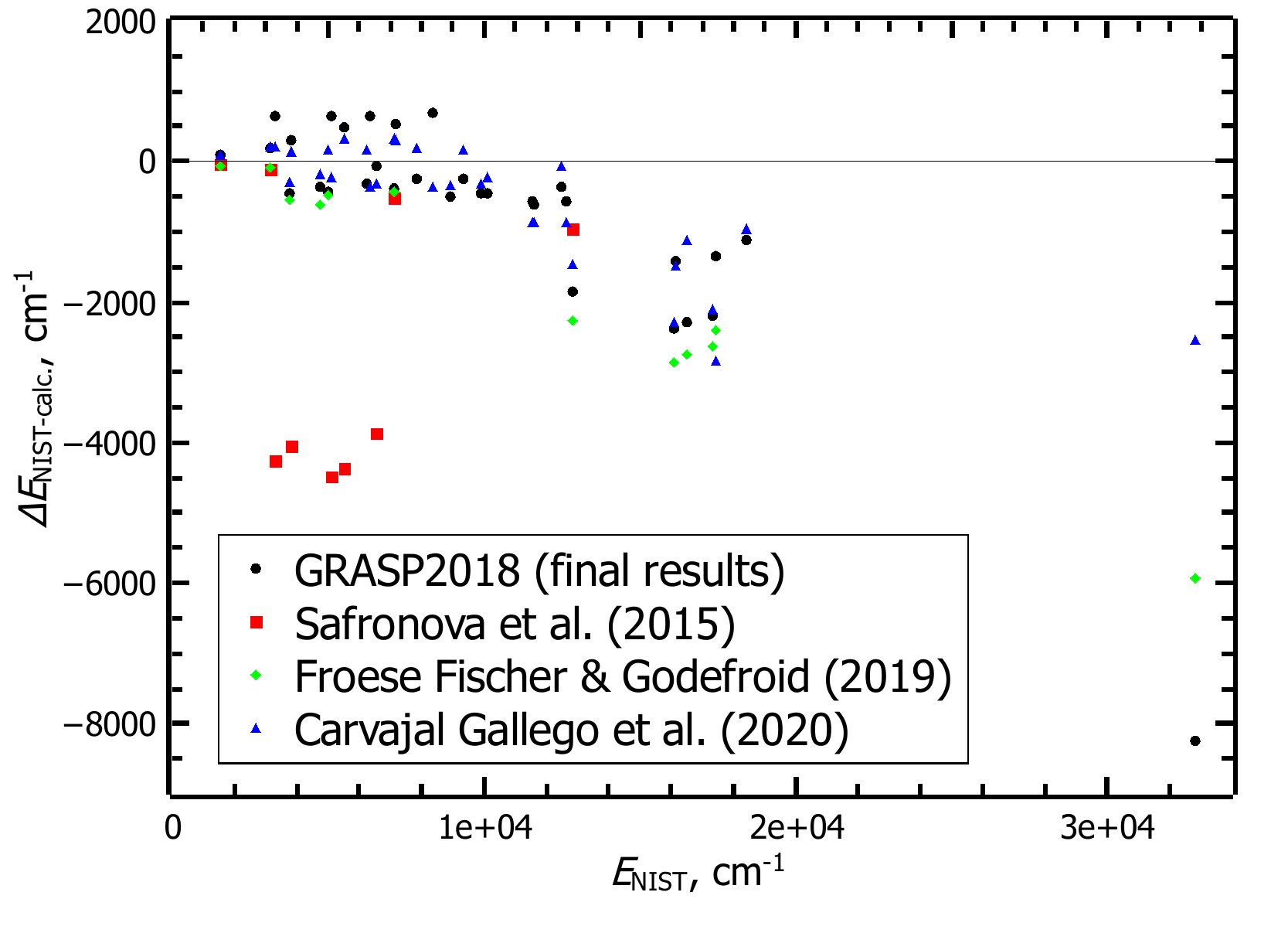}
\caption{\label{comparison_with_others} Differences between NIST ASD energy levels and those of the present {\sc Grasp}2018, and other calculations
(in cm$^{-1}$).}
\end{figure}

The final results were also compared with other theoretical calculations.
The comparison is presented in Fig. \ref{comparison_with_others}.
It should be mentioned that \citet{Safronova_2015_PRA} studied only some levels of the ground and first excited configurations, while 
\citet{Fischer_2019_PRA} studied only energy levels of the ground configuration.
The differences of other theoretical results with NIST data for most energy levels are similar. 
The rms for energy levels of the ground configuration by \citet{Fischer_2019_PRA} from the NIST data 
with excluded level (4f$^2~^1S_0$) is 1777 cm$^{-1}$. 
The rms for energy levels of the ground configuration by \citet{Gallego_2021_MNRAS} from the NIST data 
with excluded level (4f$^2~^1S_0$) is 1392 cm$^{-1}$, and 565 cm$^{-1}$ for the first excited configuration.
The levels of the first excited configurations computed by \citet{Safronova_2015_PRA} disagree about 4000 cm$^{-1}$.
The final results of the energy spectra along with the atomic state function
composition in $LS$-coupling are given in Table \ref{Ce_III_composition}.

\begin{table}
\caption{Atomic state function composition (up to three $LS$ components with a contribution > 0.02 of the total atomic state function) in $LS$-coupling and energy levels (in cm$^{-1}$) for Ce~III. Energy levels are given relative to the ground state.
}
\label{Ce_III_composition}
\centering
\begin{tabular}{rllr}
\hline\hline
No. & State & $LS$ composition & $E_{RCI}$   \\
\hline
\noalign{\smallskip}
1   & $\mathrm{4f^{2}~^{3}H_{4}}$ & 0.90                                                                                       & 0            \\
2   & $\mathrm{4f^{2}~^{3}H_{5}}$ & 0.92                                                                                       & 1421        \\
3   & $\mathrm{4f\,5d~^{1}G_{4}^{\circ}}$   & 0.66 + 0.20~$\mathrm{4f\,5d~^{3}H^{\circ}}$ + 0.04~$\mathrm{4f\,5d~^{3}F^{\circ}}$
& 2626        \\
4   & $\mathrm{4f^{2}~^{3}H_{6}}$ & 0.92                                                                                       & 2933        \\
5   & $\mathrm{4f\,5d~^{3}F_{2}^{\circ}}$   & 0.73 + 0.17~$\mathrm{4f\,5d~^{1}D^{\circ}}$                                   & 3507        \\
6   & $\mathrm{4f^{2}~^{3}F_{2}}$ & 0.90                                                                                       & 4219        \\
7   & $\mathrm{4f\,5d~^{3}H_{4}^{\circ}}$   & 0.70 + 0.16~$\mathrm{4f\,5d~^{1}G^{\circ}}$ + 0.04~$\mathrm{4f\,5d~^{3}F^{\circ}}$
& 4489        \\
8   & $\mathrm{4f\,5d~^{3}F_{3}^{\circ}}$   & 0.89                                                                                       & 5009        \\
9   & $\mathrm{4f^{2}~^{3}F_{3}}$ & 0.92                                                                                       & 5125        \\
10  & $\mathrm{4f^{2}~^{3}F_{4}}$ & 0.61 + 0.30~$\mathrm{4f^{2}~^{1}G}$                                & 5435        \\
11  & $\mathrm{4f\,5d~^{3}H_{5}^{\circ}}$   & 0.91                                                                                       & 5709        \\
12  & $\mathrm{4f\,5d~^{3}G_{3}^{\circ}}$   & 0.86 + 0.03~$\mathrm{4f\,5d~^{1}F^{\circ}}$                                   & 6578        \\
13  & $\mathrm{4f\,5d~^{3}F_{4}^{\circ}}$   & 0.81 + 0.08~$\mathrm{4f\,5d~^{1}G^{\circ}}$                                   & 6613        \\
14  & $\mathrm{4f\,5d~^{1}D_{2}^{\circ}}$   & 0.69 + 0.17~$\mathrm{4f\,5d~^{3}F^{\circ}}$ + 0.03~$\mathrm{4f\,5d~^{3}D^{\circ}}$
& 6634        \\
15  & $\mathrm{4f^{2}~^{1}G_{4}}$ & 0.60 + 0.31~$\mathrm{4f^{2}~^{3}F}$                                & 7515        \\
16  & $\mathrm{4f\,5d~^{3}H_{6}^{\circ}}$   & 0.91                                                                                       & 7666        \\
17  & $\mathrm{4f\,5d~^{3}G_{4}^{\circ}}$   & 0.89                                                                                       & 8081        \\
18  & $\mathrm{4f\,5d~^{3}D_{1}^{\circ}}$   & 0.88                                                                                       & 9423        \\
19  & $\mathrm{4f\,5d~^{3}G_{5}^{\circ}}$   & 0.90                                                                                       & 9580        \\
20  & $\mathrm{4f\,5d~^{3}D_{2}^{\circ}}$   & 0.87 + 0.03~$\mathrm{4f\,5d~^{1}D^{\circ}}$                                   & 10351       \\
21  & $\mathrm{4f\,5d~^{3}D_{3}^{\circ}}$   & 0.67 + 0.22~$\mathrm{4f\,5d~^{1}F^{\circ}}$                                   & 10579       \\
22  & $\mathrm{4f\,5d~^{3}P_{0}^{\circ}}$   & 0.90                                                                                       & 12152       \\
23  & $\mathrm{4f\,5d~^{3}P_{1}^{\circ}}$   & 0.87 + 0.03~$\mathrm{4f\,5d~^{1}P^{\circ}}$                                   & 12223       \\
24  & $\mathrm{4f\,5d~^{1}F_{3}^{\circ}}$   & 0.65 + 0.23~$\mathrm{4f\,5d~^{3}D^{\circ}}$                                   & 12862       \\
25  & $\mathrm{4f\,5d~^{3}P_{2}^{\circ}}$   & 0.88 + 0.02~$\mathrm{4f\,5d~^{1}D^{\circ}}$                                   & 13212       \\
26  & $\mathrm{4f^{2}~^{1}D_{2}}$ & 0.85 + 0.05~$\mathrm{4f^{2}~^{3}P}$                                & 14693       \\
27  & $\mathrm{4f\,5d~^{1}H_{5}^{\circ}}$   & 0.89                                                                                       & 17558       \\
28  & $\mathrm{4f^{2}~^{3}P_{0}}$ & 0.91                                                                                       & 18449       \\
29  & $\mathrm{4f^{2}~^{1}I_{6}}$ & 0.91                                                                                       & 18764       \\
30  & $\mathrm{4f^{2}~^{3}P_{1}}$ & 0.91                                                                                       & 18817       \\
31  & $\mathrm{4f^{2}~^{3}P_{2}}$ & 0.86 + 0.05~$\mathrm{4f^{2}~^{1}D}$                                & 19508       \\
32  & $\mathrm{4f\,5d~^{1}P_{1}^{\circ}}$   & 0.86 + 0.03~$\mathrm{4f\,5d~^{3}P^{\circ}}$                                   & 19570       \\
33  & $\mathrm{4f^{2}~^{1}S_{0}}$ & 0.89                                                                                  & 41076       \\
\hline
\hline
\end{tabular}
\end{table}

\subsection{Evaluation of transition data}
\label{Evaluation_transition}
\begin{figure}
\centering
\includegraphics[width=\hsize]{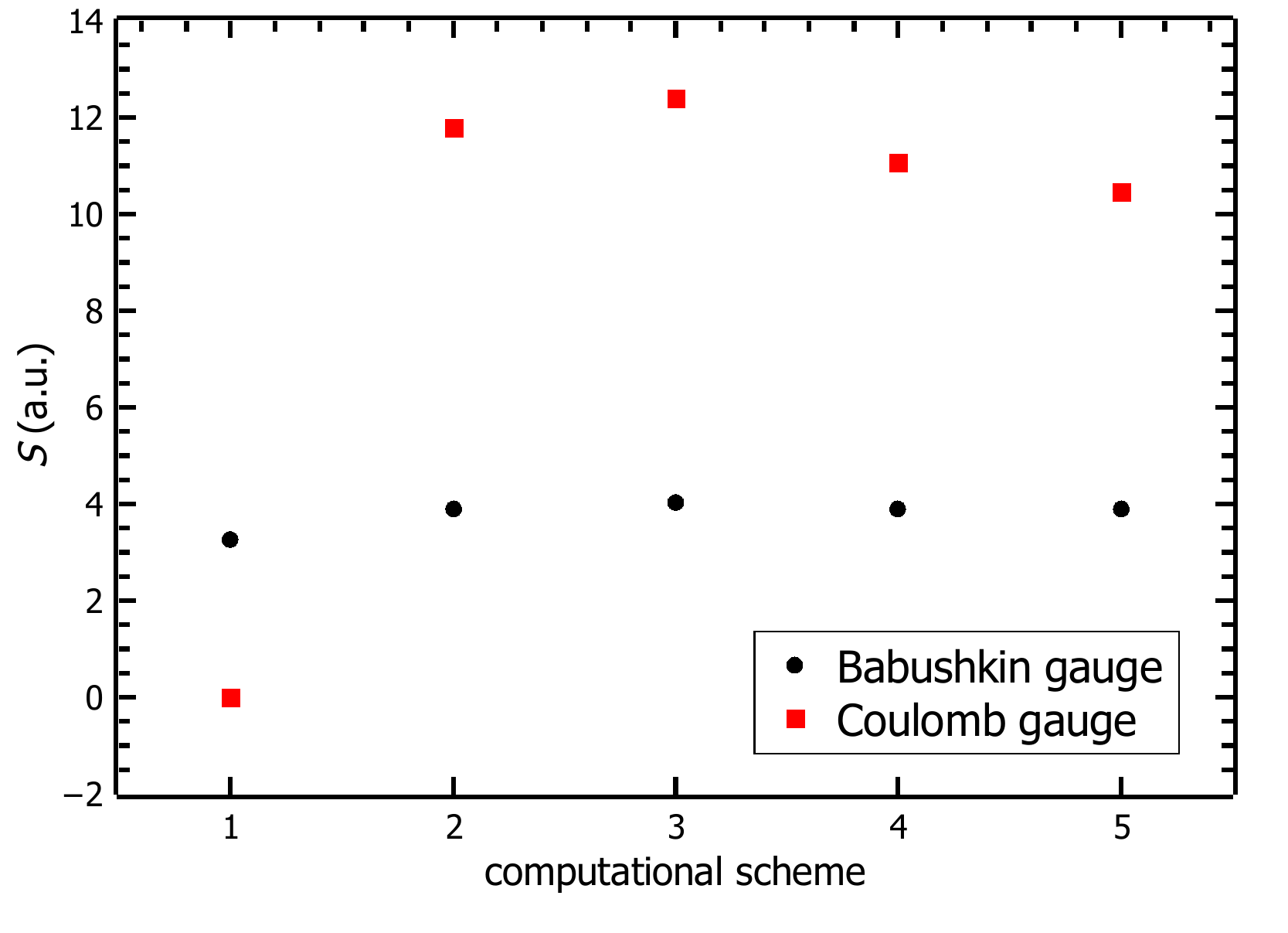}
\caption{\label{strategy_comparison_transition} Contributions of electron correlation effects to line strengths of the 
$\mathrm{4f^2~^3H_4 - 4f\,5d~^3G_3}$ transition.
On $x$ axis the computational schemes are marked: 1 - \textbf{5s5p}; 2 - \textbf{+4d}; 3 - \textbf{+4d4p};
4 - \textbf{+4dSrD4s4p}; 5 - \textbf{+4dSrD3d4s4p}.}
\end{figure}

\begin{figure}
\centering
\includegraphics[width=\hsize]{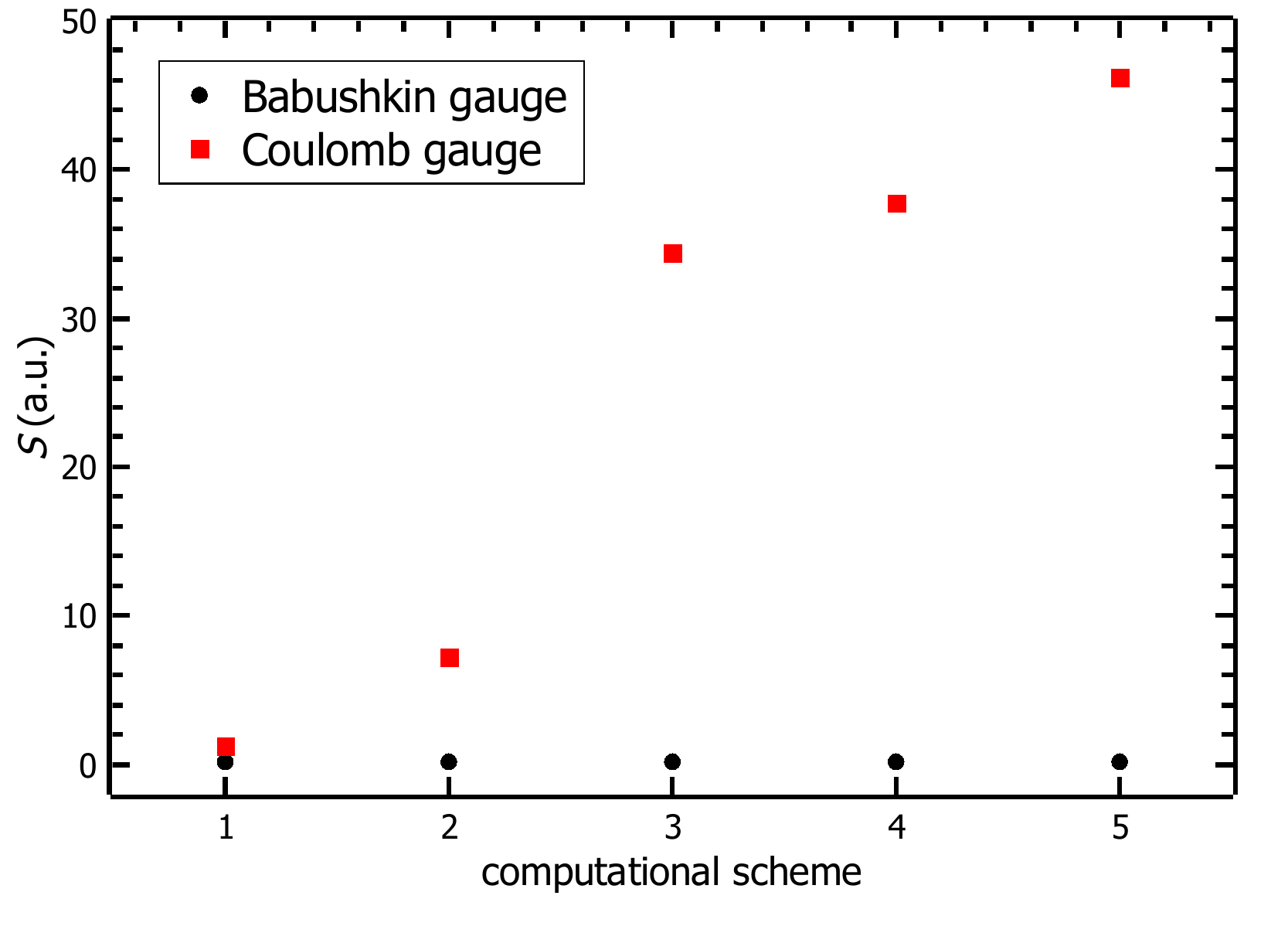}
\caption{\label{strategy_comparison_transition2} Contributions of electron correlation effects to line strengths of the 
$\mathrm{4f^2~^3F_4 - 4f\,5d~^3F_3}$ transition.
On $x$ axis the computational schemes are marked: 1 - \textbf{5s5p}; 2 - \textbf{+4d}; 3 - \textbf{+4d4p};
4 - \textbf{+4dSrD4s4p}; 5 - \textbf{+4dSrD3d4s4p}.}
\end{figure}

The importance of correlations for transition data was also studied.
As it was mentioned above, the importance of correlation effects was studied for $\mathrm{4f^2}$~$(J=4)$ and $\mathrm{4f\,5d}$~$(J=3)$ levels.
In Figs. \ref{strategy_comparison_transition} and \ref{strategy_comparison_transition2} are presented the contributions of the electron correlation effects to the line strengths for a few transitions. 
From the figures it seen that there are large disagreements between the Babushkin and Coulomb forms.
By analyzing the impact of electron correlation to line strength it seen that the Babushkin form is more stable than the Coulomb form for all studied transitions.
It is important to include correlations from 4d orbitals, and further opening the 4p and 4s core have a 
smaller effect and almost do not change by opening the 3d orbitals.
The same trend is observed for the remaining studied transitions, too.
Therefore, the \textbf{+4d SrD4s4p} scheme was chosen to calculate all E1 transitions between two configurations. 

Recently, a new method based on the stationary second-order Rayleigh-Schr\"odinger 
many-body perturbation theory in an irreducible tensorial form (RSMBPT)
has been developed to estimate various  correlations.
This method was already tested on light and moderate complexity ions, taking into account the core-valence (CV) and core (C) correlations \citep{RSMBPT_CV,RSMBPT_C} 
to compute the energy levels.
The expressions for estimating the core-core (CC) and valence-valence (VV) correlations have been derived and the papers are in preparation \citep{RSMBPT_CC,RSMBPT_VV}.
Since using the regular {\sc Grasp}2018 computational scheme (\textbf{+4d SrD4s4p}) the agreement between obtained line strengths in the 
Babushkin and Coulomb gauges is quite
poor, the RSMBPT method was applied for Ce~III calculations.
These studies were carried out only for $\mathrm{4f^2}$~$(J=4)$ and $\mathrm{4f\,5d}$~$(J=3)$ levels to 
reduce the computational resources.
In these computations we use the same wavefunctions as described above. We open the core till the 3s subshell. Thus using RSMBPT method 
we select the most important configurations 
by the CV, C, CC, VV correlations impact with the specified fraction (expressed in the percentage) of the total contribution (in this case 97\%).
Other correlations, which can not be estimated using the RSMBPT method were included in the RCI calculations in a regular way.
Further the transitions were computed between these levels. 
The results of these computations are marked as \textbf{RCI~(RSMBPT)~97\%}.
The line strengths in two gauges together with the cancellation factors (CFs), $G_{S=0}$ parameter 
and accuracy classes according to the QQE method \citep{Rynkun_2021_Ce,Gaigalas_2022_Pr} are presented in Table \ref{com_line_st_schemes} for few transitions. As seen from Table \ref{com_line_st_schemes}, by opening deeper core (up to 3s orbital), the line strength in Babushkin gauge has changed very little. 
Meanwhile the results and core correlations analysis show that the Coulomb gauge is unstable and very sensitive to correlations.

As is well known, the Coulomb gauge describes better the part of the radial wavefunctions that is closer to the nucleus.
We have therefore tried to describe better this part by including the core correlations in the MCDHF calculations \citep{atoms_Per}.
The wavefunctions were obtained in following way:
the first steps getting radial wavefunctions for 1s,...,5d orbitals are the same as described above.
The orbitals belonging to the $\mathrm{OS_{1-3}}$ were optimized using MCDHF in the extended optimal level EOL scheme.
AS was generated by allowing SD electron substitutions from the $\mathrm{4f, 5d}$ shells 
and S substitutions from $\mathrm{3s, 3p, 3d, 4s, 4p, 4d, 4s, 5p}$ 
to the OS: $\mathrm{OS_1 = \{6s,6p,6d,5f,5g\}}$, $\mathrm{OS_2 = \{7s,7p,7d,6f,6g,6h\}}$, 
$\mathrm{OS_3 = \{8s,8p,8d,7f,7g,7h\}}$.
Further using RSMBPT method 
the most important configurations 
by the CV, C, CC, VV correlations impact with the specified fraction (expressed in the percentage) of the total contribution (in this case with 97\% and 98\%) was selected and performed RCI computations including correlations, which can not be estimated using the RSMBPT method.
The transitions were computed between these levels and these results are marked as \textbf{RCI~(RSMBPT)~97\%~new wavefunctions} and \textbf{RCI~(RSMBPT)~98\%~new wavefunctions}. The results are presented in Table \ref{com_line_st_schemes}.
As seen from the table, using the RSMBPT method with recalculated wavefunctions the agreement between two forms is better. It should also be noted that the values of the line strengths are close to those of the \textbf{+4d SrD4s4p} strategy in the Babushkin gauge. Comparing the CF in both forms using different strategies, it is seen that the CF is larger in the Babushkin gauge in all cases. A small value of the CF (less than 0.1 or 0.05 \citep{Cowan}) indicates that the calculated transition parameter is affected by strong cancellation effects. The values of CF for these given transitions are CF$_B$ > 0.205, and CF$_C$ > 0.00139. We also see that CF$_C$ changes much more than CF$_B$ using different computational schemes,  indicating that the Coulomb gauge is much more sensitive to the correlations. 
So, after all the investigations and analysis, the Babushkin gauge should be more accurate and be closer to the exact value.
Using the RSMBPT method for Ce~III computations we do not get good energy differences between levels of the ground and excited configurations. Therefore, further investigations using the RSMBPT method for more complex ions are needed, as such computations are extremely demanding even using standard schemes.

\begin{table*}
{\scriptsize
\caption{Comparison of computed line strengths ($S$ in a.u.)  using different strategies. 
$S_B$ is the line strength in the Babushkin gauge, $S_C$ is the line strength in the Coulomb gauge. 
Accuracy classes (last column) match   
the NIST ASD~\citep{NIST_ASD} terminology (AA $\leq$ 1~\%, A${+}$ $\leq$ 2~\%, A $\leq$ 3~\%, 
B${+}$ $\leq$ 7~\%, B $\leq$ 10~\%, C${+}$  $\leq$ 18~\%,  C $\leq$ 25~\%,  
D${+}$ $\leq$ 40~\%, D $\leq$ 50~\%, and E $>$ 50~\%).
}
\label{com_line_st_schemes}
\centering
\begin{tabular}{l l l r r r r r r l}
\hline\hline
\multicolumn{1}{c}{Strategy} &\multicolumn{1}{c}{State even}
 & \multicolumn{1}{c}{State odd} 
& \multicolumn{1}{c}{$\lambda$} & \multicolumn{1}{c}{$S_B$} & \multicolumn{1}{c}{$S_C$} & \multicolumn{1}{c}{CF$_B$} & \multicolumn{1}{c}{CF$_C$} & \multicolumn{1}{c}{$G_{S=0}$} & \multicolumn{1}{c}{Acc.}\\
\hline
\noalign{\smallskip}
\textbf{+4d SrD4s4p}	&$\mathrm{4f^2~^3H_4}	$  &$\mathrm{4f\,5d~^3G^o_3}	$&15202.69& 3.90& 11.1\;\,~& 0.268& 0.00263& 3.48& E \\
\textbf{RCI~(RSMBPT)~97\%} &$\mathrm{4f^2~^3H_4}	$  &$\mathrm{4f\,5d~^3G^o_3}	$&15541.07& 4.14& 25.4\;\,~& 0.332& 0.00705& 2.37& E \\
new wavefunctions\\
\textbf{RCI~(RSMBPT)~97\%} &$\mathrm{4f^2~^3H_4}	$  &$\mathrm{4f\,5d~^3G^o_3}	$&6766.95& 4.32& 4.85& 0.380& 0.00740& 25.2\;~& C+ \\
\textbf{RCI~(RSMBPT)~97\%} &$\mathrm{4f^2~^3H_4}	$  &$\mathrm{4f\,5d~^3G^o_3}	$&6410.19& 4.37& 4.52& 0.380& 0.00724& 81.6\;~& B+ \\
\\
\\
\textbf{+4d SrD4s4p}	&$\mathrm{4f^2~^3F_4}	$  &$\mathrm{4f\,5d~^3D^o_3}	$&19440.90& 2.58& 9.02& 0.205& 0.00139& 3.04 & E \\
\textbf{RCI~(RSMBPT)~97\%} &$\mathrm{4f^2~^3F_4}	$  &$\mathrm{4f\,5d~^3D^o_3}	$&16128.81& 2.63& 14.0\;\,~& 0.245& 0.00395& 2.49 & E \\
new wavefunctions\\
\textbf{RCI~(RSMBPT)~97\%} &$\mathrm{4f^2~^3F_4}	$  &$\mathrm{4f\,5d~^3D^o_3}	$&6987.76& 2.71& 3.02& 0.283& 0.00496& 27.4\;~& C+ \\
\textbf{RCI~(RSMBPT)~98\%} &$\mathrm{4f^2~^3F_4}	$  &$\mathrm{4f\,5d~^3D^o_3}	$&6726.20& 2.77& 3.46& 0.288& 0.00555& 13.4\;~& C \\
\\
\\
\textbf{+4d SrD4s4p}	&$\mathrm{4f^2~^1G_4}	$  &$\mathrm{4f\,5d~^1F^o_3}	$&18701.34& 3.12& 10.7\;\,~& 0.263& 0.00197& 3.08& E \\
\textbf{RCI~(RSMBPT)~97\%} &$\mathrm{4f^2~^1G_4}	$  &$\mathrm{4f\,5d~^1F^o_3}	$&20343.89& 3.23& 26.7\;\,~& 0.329& 0.00530& 2.17& E \\
new wavefunctions\\
\textbf{RCI~(RSMBPT)~97\%} &$\mathrm{4f^2~^1G_4}	$  &$\mathrm{4f\,5d~^1F^o_3}	$&7652.41& 3.29& 4.38& 0.378& 0.00666& 10.6\;~& D+ \\
\textbf{RCI~(RSMBPT)~98\%} &$\mathrm{4f^2~^1G_4}	$  &$\mathrm{4f\,5d~^1F^o_3}	$&7114.95& 3.33& 4.24& 0.379& 0.00697& 12.4\;~& D+ \\

\hline
\hline
\end{tabular}
}
\end{table*}

Since no experimental transition data for the studied transitions of 
Ce$^{2+}$ are available, the obtained data are compared with other theoretical results.
A comparison of the computed line strengths from the present work and
the results of \citet{Gallego_2021_MNRAS}, \citet{Biemont_2002_MNRAS}, 
\citet{Wyart_1998}, and \citet{Tanaka20} \citep[see also][]{Domoto_2022} is given in Table \ref{comparison_line_st}.
There are disagreements between the theoretical results.
Comparing the line strengths computed in this work 
with the results by \citet{Gallego_2021_MNRAS}, \citet{Biemont_2002_MNRAS}, 
\citet{Wyart_1998}, it is seen that they are in better agreement with the Babushkin form.
Analyzing the ratios between the Babushkin and the Coulomb gauges given by \citet{Gallego_2021_MNRAS},
it is seen the large disagreement between two gauges in their results, too.

\begin{table*}
{\scriptsize
\caption{Comparison of computed line strengths ($S$ in a.u.) with results from \citet{Gallego_2021_MNRAS,Biemont_2002_MNRAS,Wyart_1998,Tanaka20}.
$S_B$ is the line strength in the Babushkin gauge, $S_C$ is the line strength in the Coulomb gauge. $B$/$C$ is the ratio Babushkin over Coulomb gauges.
}
\label{comparison_line_st}
\centering
\begin{tabular}{l l r r cc c c c c}
\hline\hline
\multicolumn{1}{c}{\multirow{2}{0.86cm}{State even}}
 & \multicolumn{1}{c}{\multirow{2}{0.80cm}{State odd}} 
&\multicolumn{2}{c}{Present}&&
\multicolumn{2}{c}{\cite{Gallego_2021_MNRAS}} & \multicolumn{1}{c}{\cite{Biemont_2002_MNRAS}}&\multicolumn{1}{c}{\cite{Wyart_1998}}&\multicolumn{1}{c}{\cite{Tanaka20}} \\
\cline{3-4} \cline{6-7} 
 &  &  \multicolumn{1}{c}{$S_B$}&\multicolumn{1}{c}{$S_C$} &&
\multicolumn{1}{c}{$S$} & \multicolumn{1}{c}{$B$/$C$} & \multicolumn{1}{c}{$S$ (CF)}&\multicolumn{1}{c}{$S$}&\multicolumn{1}{c}{$S$} \\
\hline
\noalign{\smallskip}
$\mathrm{4f^2~^3H_4}	$  &$\mathrm{4f\,5d~^3F^o_3}	$&   3.108E$-$02 & 9.391E$-$02 && 7.637E$-$02 &8.46E$-$02 &&& 1.227E$-$03 \\
$\mathrm{4f^2~^3H_4}	$  &$\mathrm{4f\,5d~^3G^o_3}	$&   3.899E+00   & 1.106E+01   && 2.767E+00   &1.10E$-$01 &&6.230E+00& 7.313E+00  \\
$\mathrm{4f^2~^3H_4}	$  &$\mathrm{4f\,5d~^3D^o_3}	$&   1.150E$-$02 & 2.417E$-$02 && 4.291E$-$03 &4.57E$-$01 &1.324E$-$02 (0.056)&& 2.086E$-$02 \\
$\mathrm{4f^2~^3H_4}	$  &$\mathrm{4f\,5d~^1F^o_3}	$&   7.346E$-$03 & 1.401E$-$02 && 1.580E$-$03 &1.66E$-$01 &6.613E$-$03 (0.019)&1.053E$-$02 &9.677E$-$03  \\
$\mathrm{4f^2~^3F_4}	$  &$\mathrm{4f\,5d~^3F^o_3}	$&   1.757E$-$01 & 3.773E+01   && 9.274E$-$02 &6.53E$-$04 &&& 8.591E$-$02 \\
$\mathrm{4f^2~^3F_4}	$  &$\mathrm{4f\,5d~^3G^o_3}	$&   3.331E$-$05 & 1.748E$-$02 && 2.740E$-$03 &4.14E$-$03 &&& 2.194E$-$03 \\
$\mathrm{4f^2~^3F_4}	$  &$\mathrm{4f\,5d~^3D^o_3}	$&   2.585E+00   & 9.026E+00   && 1.935E+00   &1.20E$-$01 &&4.056E+00& 8.588E$-$02  \\
$\mathrm{4f^2~^3F_4}	$  &$\mathrm{4f\,5d~^1F^o_3}	$&   6.923E$-$02 & 1.674E$-$01 && 2.513E$-$01 &1.62E$-$01 &&& 4.986E+00   \\
$\mathrm{4f^2~^1G_4}	$  &$\mathrm{4f\,5d~^3F^o_3}	$&   1.515E$-$02 & 5.870E$-$02 && 2.905E$-$02 &1.81E$-$02 &&& 2.340E$-$01 \\
$\mathrm{4f^2~^1G_4}	$  &$\mathrm{4f\,5d~^3G^o_3}	$&   3.637E$-$03 & 7.903E$-$03 && 2.107E$-$03 &9.92E$-$03 &&& 2.153E$-$03 \\
$\mathrm{4f^2~^1G_4}	$  &$\mathrm{4f\,5d~^3D^o_3}	$&   2.641E$-$05 & 3.506E$-$04 && 5.961E$-$02 &1.04E$-$01 &&& 4.680E+00   \\
$\mathrm{4f^2~^1G_4}	$  &$\mathrm{4f\,5d~^1F^o_3}	$&   3.122E+00   & 1.066E+01   && 2.118E+00   &1.29E$-$01 &&4.814E+00& 4.759E$-$01 \\
\hline
\hline
\end{tabular}
}
\end{table*}

\begin{figure*}
\centering
\includegraphics[width=\hsize]{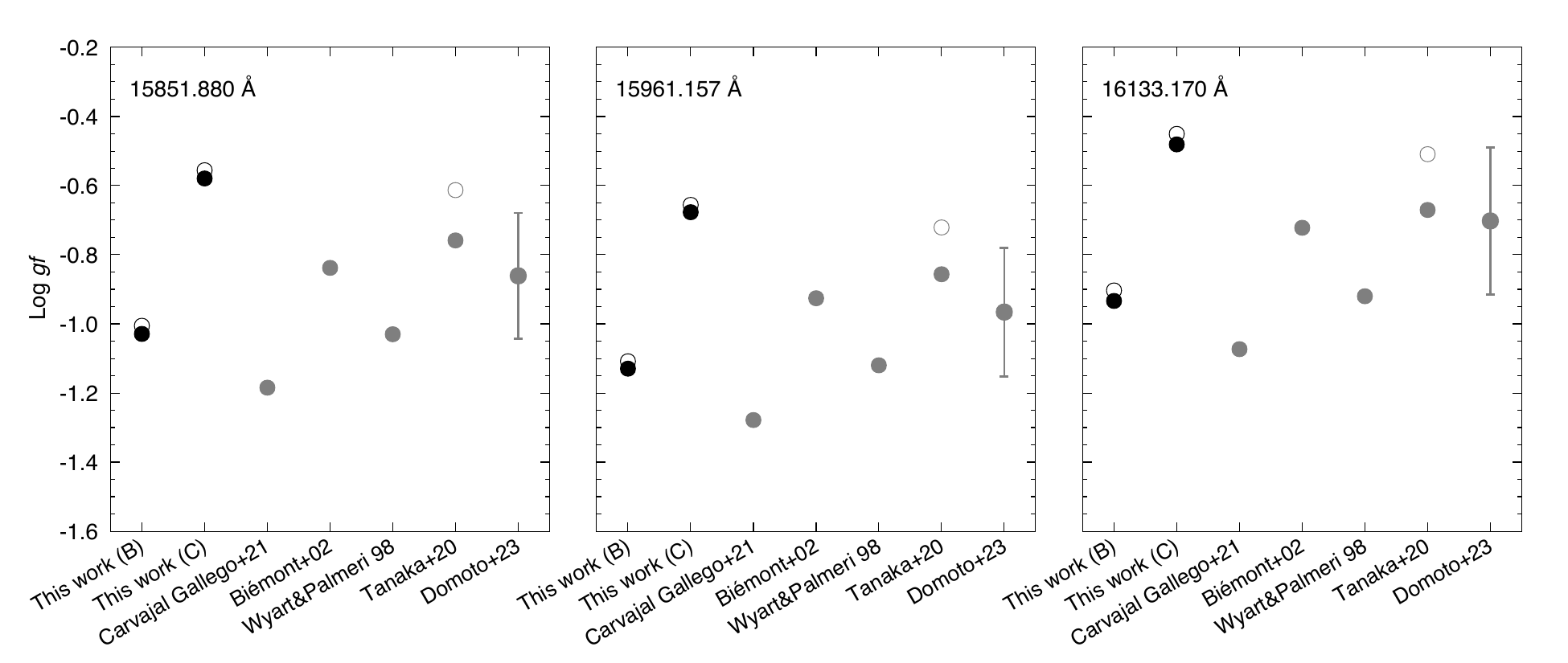}
\caption{\label{fig:gf}
Comparison of the computed $gf$-values (black) with results from \citet{Gallego_2021_MNRAS}, \citet{Biemont_2002_MNRAS}, \citet{Wyart_1998}, \citet{Tanaka20}, and \citet{Domoto_2023} (gray).
For the computed values in this work, the results using \textbf{+4d SrD4s4p} scheme as the final values are shown.
B and C indicate the results in the Babushkin and Coulomb gauges, respectively.
Note that the filled symbols show the calibrated values based on the line strengths, considering the differences in theoretical energy levels to experimental energy levels, while the open symbols show the values as the results are.}
\end{figure*}

A comparison of the computed $gf$-values and other theoretical results is also shown in Fig. \ref{fig:gf}.
Here, we show only the $gf$-values of the three transitions (whose wavelengths are shown in each panel) that are the strongest and important to interpret the spectral features of kilonovae \citep[see Section \ref{sec:spec}]{Domoto_2022}.
The ``astrophysical $gf$-values'' \citep{Domoto_2023} are also shown in the figure.
The final results in the Babushkin gauge are closer to the values from other works.

The final results (using the \textbf{+4d SrD4s4p} scheme) of the transition data,
as wavelengths, weighted oscillator strengths, line strengths, and
transition rates of the E1 transitions, are given in Table \ref{Ce_III_transition_data}
 along with the $G_{S=0}$ parameter, and the estimated accuracy for line strengths in the Babushkin gauge according to the QQE method \citep{Rynkun_2021_Ce,Gaigalas_2022_Pr}. The uncertainties of the line strengths were estimated using the QQE method, since experimental data and other theoretical results with estimated error bars are not available. The full table is available as online supplementary material.

\begin{table*}
{\scriptsize
\caption{Transition wavelengths $\lambda$ (in \AA), transition rates $A$ (in s$^{-1}$), weighted oscillator strengths $gf$, and line strengths $S$ (in a.u.) for E1 transitions of the Ce~III [The full table is available online as supplementary material].  Accuracy classes (last column) match   
the NIST ASD~\citep{NIST_ASD} terminology (AA $\leq$ 1~\%, A${+}$ $\leq$ 2~\%, A $\leq$ 3~\%, 
B${+}$ $\leq$ 7~\%, B $\leq$ 10~\%, C${+}$  $\leq$ 18~\%,  C $\leq$ 25~\%,  
D${+}$ $\leq$ 40~\%, D $\leq$ 50~\%, and E $>$ 50~\%).}
\label{Ce_III_transition_data}
\centering
\begin{tabular}{ccccccccccccl}
\hline\hline
Nl& Nu& $\lambda$& $A_B$ & $gf_B$ & $S_B$ & $A_C$ & $gf_C$ & $S_C$ & CF$_B$ & CF$_C$ & $G_{S=0}$ & Acc.\\
\hline
\noalign{\smallskip}
    1&    3&  3.808421E+04&  1.782E+03&  3.487E$-$03&  4.372E$-$01&  2.084E+04&  4.079E$-$02&  5.114E+00  & 5.58E-02& 2.73E-04& 1.99850E+00 & E  \\
    1&    7&  2.227851E+04&  1.219E+04&  8.162E$-$03&  5.986E$-$01&  6.657E+04&  4.458E$-$02&  3.270E+00  & 8.36E-02& 5.94E-04& 2.47194E+00 & E  \\
    1&    8&  1.996454E+04&  1.130E+03&  4.728E$-$04&  3.108E$-$02&  3.416E+03&  1.429E$-$03&  9.391E$-$02& 5.32E-03& 3.41E-05& 3.32942E+00 & E  \\
    1&   11&  1.751691E+04&  1.419E+03&  7.179E$-$04&  4.140E$-$02&  5.839E+03&  2.955E$-$03&  1.704E$-$01& 3.93E-02& 1.53E-04& 2.78906E+00 & E  \\
    1&   12&  1.520270E+04&  3.212E+05&  7.791E$-$02&  3.899E+00  &  9.109E+05&  2.209E$-$01&  1.106E+01  & 2.68E-01& 2.63E-03& 3.48187E+00 & E  \\
\hline
\hline
\end{tabular}
}
\end{table*}

\section{Applications to kilonova spectra}
\label{sec:spec}
Here, we apply the final results of $gf$-values of the Ce III lines to calculate kilonova spectra.
To calculate synthetic spectra of kilonovae, we use a wavelength-dependent radiative transfer simulation code \citep{Tanaka13, Tanaka14, Tanaka17, Tanaka18, Kawaguchi18, Kawaguchi20}.
The photon transfer is calculated by the Monte Carlo method.
The setup of simulation is identical to that in \citet{Domoto_2022} and \citet{Domoto_2023}, but we adopt the $gf$-values of the Ce III lines computed in this work (Table \ref{Ce_III_transition_data}).
For more details of the simulation, we refer the readers to \citet{Domoto_2022}.

Among the transitions between the states of the ground configuration and the first excited configuration, it has been shown that the three transitions shown in Fig. \ref{fig:gf} gives the strongest contribution to the NIR spectral features of kilonovae \citep{Domoto_2022}.
Thus, although we have all the transition probabilities of the computed transitions, we take the $gf$-values of these three lines to see their effects on the spectral features.
As the line strengths suggest that the Babushkin gauge is more reliable, we adopt the $gf$-values in the Babushkin gauge in the simulations (i.e., the 5th column of Table \ref{Ce_III_transition_data}).

\begin{figure}
\centering
\includegraphics[width=\hsize]{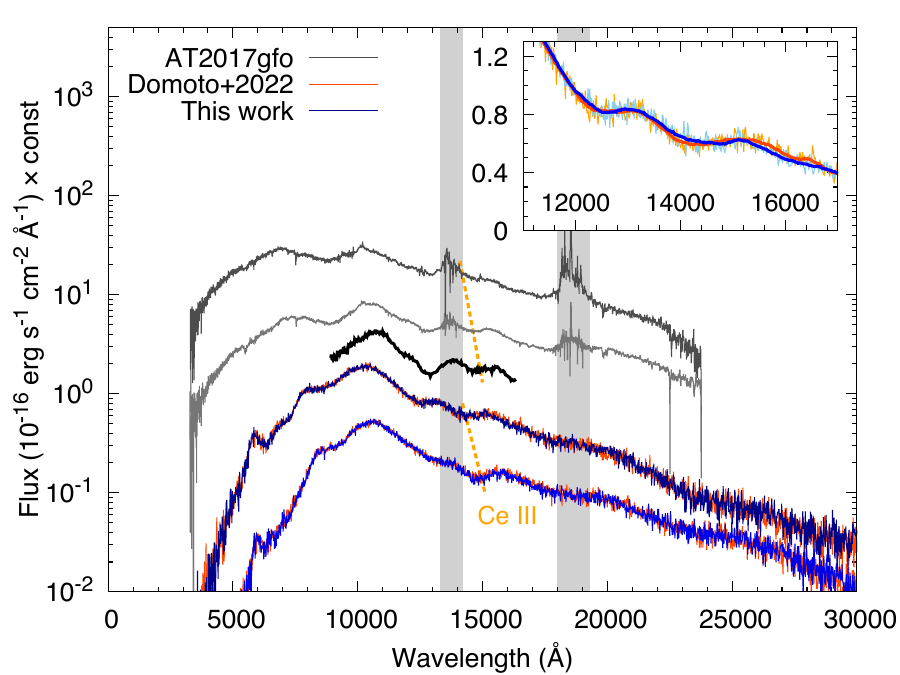}
\caption{\label{fig:spec}
Comparison between the synthetic spectra using the final $gf$-values computed in the Babushkin gauge (blue) and the observed spectra of AT2017gfo taken with VLT \citep[gray,][]{Pian2017, Smartt2017} at $t=2.5$ and 3.5 days after the merger as well as with HST \citep[black,][]{Tanvir2017} at $t=4.5$ days after the merger.
The orange lines are the synthetic spectra of \citet{Domoto_2022}.
Gray shaded areas show the regions of strong atmospheric absorption.
Spectra are vertically shifted for visualization.
The inset shows the enlarged view of the synthetic spectra in the NIR region at $t=2.5$ days.
The curves in light colors show the original results, while those in dark colors show smoothed spectra for visualization.
}
\end{figure}

Fig. \ref{fig:spec} shows the comparison between the synthetic spectra (blue) and the observed spectra of AT2017gfo taken with the Very Large Telescope (VLT) at $t=1.5$, 2.5, and 3.5 days after the merger \citep[gray,][]{Pian2017, Smartt2017}.
The observed spectra at $t=4.5$ days after the merger taken with the Hubble Space Telescope (HST), which are not affected by telluric absorption, is also shown \citep[black,][]{Tanvir2017}.
In the synthetic spectra, 
the absorption features appear around 14500 {\AA}, which 
 are caused by the Ce III lines.
 These lines are blueshifted according to the velocity of the line-forming region at the NIR wavelengths (e.g., $v\sim0.1\ c$ at $t=2.5$ days).
Since our calculated $gf$-values are smaller than those adopted in \citet{Domoto_2022}, the absorption features in the new synthetic spectra (orange) become slightly weaker (see the inset of Fig. \ref{fig:spec} for the enlarged view at $t=2.5$ days). 
Nevertheless, the absorption features are still clearly present, and the strength of the feature is broadly consistent with the observed spectra of AT2017gfo.
This gives the further support to the identification of the Ce III lines in the NIR spectra of kilonova.


\section{Summary and conclusions}
\label{sec:conclusions}
We calculated the energy levels of the ground and first excited configurations for the Ce III using the {\sc Grasp}2018 code.
The energy differences between the final {\sc Grasp}2018 results and the NIST ASD for two configurations up to 12000 cm$^{-1}$ reach 600 cm$^{-1}$.
The disagreement for other energies reaches 2500 cm$^{-1}$, and the largest difference (8200 cm$^{-1}$) is for the level of the ground configuration (4f$^2~^1S_0$).
The rms deviations obtained for the energy levels of the ground configuration from the NIST data are 2732 cm$^{-1}$, but excluding the level with the largest discrepancy (4f$^2~^1S_0$), the rms is 1404 cm$^{-1}$. The rms for the first excited configuration is 618 cm$^{-1}$.

We also computed E1 transition data between the levels of the ground and first excited configurations. 
The uncertainties of the E1 line strengths in the Babushkin gauge are estimated based on the QQE method
described in \cite{Rynkun_2021_Ce,Gaigalas_2022_Pr} giving the accuracy classes
according to the NIST ASD \citep{NIST_ASD}. 
The line strengths were also compared with the results of other calculations.
The analysis of the line strengths shows that the Babushkin gauge should be the more accurate, 
so we suggest use the Babushkin gauge with the assigned accuracy for the obtained results although the line strengths are assigned with E accuracy class. 
  
Finally, we performed radiative transfer simulations for kilonova spectra by using the calculated $gf$ values.
The synthetic spectra clearly show the absorption features around 14,500 \AA, which is caused by the blueshifted Ce III lines.
Therefore, our {\it ab-initio} atomic calculations support the identification of the Ce III lines in the NIR spectra of kilonova.



\section*{Acknowledgments}
This project has received funding from the 
Research Council of Lithuania (LMTLT), agreement No S-LJB-23-1
and JSPS Bilateral Joint Research Project (JPJSBP120234201).

\section*{DATA AVAILABILITY}
The data (Table \ref{Ce_III_transition_data}) underlying this article are available in
the article and in its online supplementary material.


\label{lastpage}

\end{document}